\documentclass[useAMS]{mn2e}
\usepackage{psfig}
\newif\ifAMStwofonts

\def\aap{A\&A}

\def\be{\begin{eqnarray}}
\def\ee{\end{eqnarray}}
\def\beq{\begin{equation}}
\def\eeq{\end{equation}}

\def\etal{{\it et al.}}

\def\HI{{\hbox{H~$\scriptstyle\rm I\ $}}}
\def\HII{\hbox{H~$\scriptstyle\rm II\ $}}
\def\HeI{\hbox{He~$\scriptstyle\rm I\ $}}
\def\HeII{\hbox{He~$\scriptstyle\rm II\ $}}
\def\HeIII{\hbox{He~$\scriptstyle\rm III\ $}}

\def\CR{{\tt CRASH}}
\def\crf{{\tt CRASH1}}
\def\crn{{\tt CRASH2}}

\def\lesssim{\mathrel{\hbox{\rlap{\hbox{\lower4pt\hbox{$\sim$}}}\hbox{$<$}}}}
\def\gtrsim{\mathrel{\hbox{\rlap{\hbox{\lower4pt\hbox{$\sim$}}}\hbox{$>$}}}}

\def\gtsima{$\; \buildrel \over \sim \;$}
\def\ltsima{$\; \buildrel < \over \sim \;$}
\def\prosima{$\; \buildrel \propto \over \sim \;$}
\def\gsim{\lower.5ex\hbox{\gtsima}}
\def\lsim{\lower.5ex\hbox{\ltsima}}
\def\simgt{\lower.5ex\hbox{\gtsima}}
\def\simlt{\lower.5ex\hbox{\ltsima}}

\def\simpr{\lower.5ex\hbox{\prosima}}
\def\la{\lsim}
\def\ga{\gsim}

\def\ie{{\frenchspacing\it et al. }}
\def\ie{{\frenchspacing\it i.e. }}
\def\eg{{\frenchspacing\it e.g. }}

\def\be{\begin{eqnarray}}
\def\ee{\end{eqnarray}}

\title[CRASH2: colored packets and other updates]
{CRASH2: colored packets and other updates}

\author[A. Maselli, B. Ciardi \& A. Kanekar]{A. Maselli$^{1}$, B. Ciardi$^{1}$ \& A. Kanekar$^{1,2}$\\
$^1$ Max-Planck-Institut f\"ur Astrophysik, Karl-Schwarzschild-Strasse1,
85748 Garching, Germany\\
$^2$ Indian Institute of Technology, Address, Mumbay, India
}

\pagerange{\pageref{firstpage}--\pageref{lastpage}} \pubyear{2008}
\begin{document}

\maketitle
\label{firstpage}

\begin{abstract}
In this paper we report on the improvements implemented in the
cosmological radiative transfer code \CR.
In particular we present a new multifrequency algorithm
for spectra sampling which makes use of {\it colored} photon packets:
we discuss the need for the multi-frequency approach, describe its
implementation and present the improved \CR~ performance in
reproducing the effects of ionizing radiation with an arbitrary
spectrum.
We further discuss minor changes in the code implementation which
allow for more efficient performance and an increased precision.
\end{abstract}

\begin{keywords}
 cosmology: theory - radiative transfer - methods: numerical -
 intergalactic medium
\end{keywords}

\section{Introduction}

The proper and accurate treatment of radiative transfer (RT) in
cosmological numerical simulations is a long standing problem, which
has received great and ever increasing attention in the last decades.
Despite this, the progress done in numerical RT has been somewhat
slower with respect to other fields of numerical cosmology and
astrophysics, like Large Scale Structure simulations or Magneto Hydrodynamics
schemes, which in the last years underwent an impressive series of successes
and refinements.
The reason for the lag above is mainly due to the extreme complexity
underlying the process of radiative transfer, which in a cosmological
setting couples also with gas dynamics, cosmic expansion, structure
formation and chemistry.

The seven dimension cosmological RT equation does not allow an
exact solution, neither analytic nor numerical, and the need to deal
with arbitrary geometries and a huge dynamical range in the optical
depth distribution prevents the adoption of simplifying assumptions to
reduce the dimensionality of the equation and the complexity of the
problem.
Several methods have been proposed in the literature,
relying on a variety of assumptions and approximations, and
often optimizing the treatment of RT for different
problems, namely different astrophysical configurations.
For a detailed review of the various methods developed in the literature
we refer the reader to Iliev \etal (2006; I06), where a systematic comparison
of 11 cosmological radiative transfer codes through a set of tests is
described. Since the publication of the paper, more RT codes designed for
cosmological applications have appeared in the literature (Altay, Croft \& Pelupessy
2008; Pawlik \& Schaye 2008; Finlator, \"Ozel \& Dav\'e 2008).

The strategy adopted by our group to face the problem of cosmological
radiative transfer is based on a ray-tracing Monte Carlo (MC) scheme
which applies to 3D Cartesian grids and which is based on
the particle nature of the radiation field.
This approach offers several advantages.
The radiation is described in terms of photons which, grouped
for computational convenience into monochromatic photon packets,
travel through the simulation volume along rays.
This allows to solve the transfer in one dimension: {\it (i)} by following
monochromatic packets of photons along rays one gets rid of the explicit dependence on direction
and frequency, and {\it (ii)} by casting rays into the assigned grid, one
can furthermore remove the dependency on the position. Instead of
solving directly for the intensity of the radiation field,
$I_\nu(\vec{r},\Omega,t)$, one just needs then to model the
interaction of the photons with the gas inside a cell.
This approach allows to enormously simplify the treatment of radiative
transfer, and to leave the geometry of the problem completely
arbitrary. On the other hand, as drawback, ray tracing techniques are
usually computationally expensive and tend to suffer from low angular
resolution.
These problems are partially overcome by adopting a statistical
description of the radiation field, by Monte Carlo
sampling directions and spectral energy distributions.
Differently from short and long characteristic methods, in which the
rays are cast along fixed directions, in MC ray tracing the rays are shot in random
directions, allowing for a more efficient implementation of multiple sources
and a diffuse component of the ionizing radiation, \eg recombination radiation or
the ultraviolet background (UVB), as well as anisotropic angular
emissivities. These advantages come with
the introduction of numerical noise, intrinsic to all MC schemes,
which can be nevertheless kept arbitrarily low by
increasing the number of rays shot. In practice, one always needs
to look for the best compromise with computational efficiency.

All the above ingredients have been successfully implemented in our
code \CR~ which is to date one of the main references among RT numerical
schemes used in cosmology.
The first version of the code (Ciardi \etal 2001, C01; in the following
we will refer to this version as \CR0)
integrated the principles of ray-tracing and MC to follow the
evolution of hydrogen ionization for multiple sources under the
assumption of a fixed temperature for the ionized gas.
In its actual reference version (Maselli, Ferrara \& Ciardi 2003,
MFC03; Maselli \& Ferrara 2005; in the following we will refer
to this version as \crf), the code has been further
developed by including a variety of additional physical ingredients
(like helium chemistry, temperature evolution, background radiation), and a more
sophisticated algorithm for processes like photon absorption,
reemission, as well as for the chemistry solver.
All these new ingredients make the code extremely versatile in term of
problems that can be studied.

One of the main advantages of the \CR~ implementation consists in the
straightforward way in which the matter/radiation interactions are
modeled and calculated. This allows to easily implement new
physics in the code without the need for major changes in its structure.
Illustrative examples are given by {\tt MCLy$\alpha$}
(Verhamme, Schraerer \& Maselli 2006) and  {\tt CRASH}$\alpha$ (Pierleoni,
Maselli \& Ciardi, 2007). Adopting the \CR~ algorithms for ray
tracing, spectra sampling, photon propagation and optical depth
sampling (as described in Ciardi \etal 2001),
we have implemented the physics of
line resonant scattering in {\tt MCLy$\alpha$}, a tool
mainly dedicated to the study and modeling of Ly$\alpha$ Emitters (LAEs).
Moving forward in this direction, we extended \CR~ and
implemented {\tt CRASH}$\alpha$, the single RT code to date which can
self-consistently account for the simultaneous and coupled transfer of line
and ionizing continuum radiation. This makes of {\tt CRASH}$\alpha$ a unique
tool for studying the implications of high redshift LAE surveys on
cosmic reionization and 21~cm line from neutral hydrogen in the
primordial universe.
More recently, we have started including molecular hydrogen in \CR~ to
study the inhomogeneous H$_2$ photo-dissociation occurring at the
dawn of the universe or in star forming regions. Once again, this requires
very little manipulation on the main core of the code and can be done
by complementing the existing numerical scheme with algorithms which model the new
physics to be implemented.

As the code is continuously being updated and optimized, in this paper
we describe some major changes that have been introduced in the
implementation described in MFC03, together with some test results
clearly showing the improvements achieved in performance and accuracy.

The rest of the paper is organized as follows. We first give a brief
review on the basics of the \CR~ implementation in Sec. 2. In Sec. 3 we
describe the updates done in the implementation and the new
algorithms introduced, presenting the results from test runs
designed to show the improvements obtained. Finally we summarize our
work in Sec. 4.

\section{CRASH: a Brief Description}
In this section we give a very brief and general description of the
\CR~ code. A more detailed and complete report is given
in MFC03 and also in Maselli \& Ferrara (2005), where a refined and
improved version of the implementation of the background radiation is
described. We refer the interested reader to these works for more
details on the implementation.

As discussed in the introduction, \CR~ is a radiative transfer
numerical scheme which applies to 3D Cartesian grids and which is
based on MC ray tracing techniques that are used to sample the
probability distribution
functions (PDFs) of several quantities involved in the calculation, {\it e.g.}
spectrum of the sources, emission direction, optical depth.
The algorithm follows the propagation of the ionizing radiation
through an arbitrary H/He static density field and, at the same time,
computes {\it (i)} the evolution of the thermal and
ionization state of the gas and {\it (ii)} the distortion of the ionizing
radiation field due to matter-radiation interactions, which produce the
typical RT effects of filtering, shadowing and self-shielding.
The radiation field is described in terms of photons which, for
computational convenience, are grouped into photon packets and shot
through the box. Both multiple point sources, located arbitrarily in the box,
and diffuse radiation fields ({\it e.g.} the ultraviolet background or the
radiation produced by H/He recombinations) are treated.

The energy emitted by each ionizing source is discretized into
photon packets emitted at regularly spaced time intervals.
More specifically, the total energy radiated by a single source of luminosity
$L_s$, during the total simulation time, $t_{sim}$, is
$E_s=\int_0^{t_{sim}}L_s(t_s)dt_s$. For each source, $E_s$ is distributed
in $N_p$ photon packets, emitted at the source location. In this
way, the total number of photon packets emitted by each source,
$N_p$, is the main control parameter which sets both the time and the
spatial resolution of a given run.
The emission direction of each photon packet is assigned by MC sampling
the angular PDF characteristic of the source.

The propagation of the packet through the given density field is then
followed and the impact of radiation-matter interaction on the gas
properties is computed on the fly.
Each time a photon packet pierces a cell $k$, the cell optical depth to 
ionizing continuum radiation, $\tau_c^k$, is estimated summing up the
contribution of the different absorbers (\HI, \HeI, \HeII).
As the probability for a single photon to be absorbed in such a cell is:
\begin{equation}
\label{tauPDF}
P(\tau_c^k)=1-e^{-\tau_c^k},
\end{equation}
the number of photons absorbed in the cell $k$ is the fraction 
$P(\tau_c^k)$ of packet content when it enters the cell.
The total number of photons deposited in the cell is then distributed
among the various absorbers according to the corresponding
contributions given to the total cell optical depth. From these quantities
the discrete increments of the ionization fractions and temperature
are calculated:
$\Delta x_{\rm HI}$, $\Delta x_{\rm HeI}$, $\Delta x_{\rm HeII}$,
$\Delta T$ and $(\Delta T)_n$, where $\Delta T$ is the
positive increment in temperature due to photo-heating and
$(\Delta T)_n$ the negative one associated to the increased number of
free particles.
The trajectory of the photon packet is followed until its photon content is
extinguished or, if periodic boundary conditions are not assumed,
until it exits the simulation volume.

The time evolution of the gas physical properties (ionization
fractions and temperature) is computed solving in each cell
the appropriate set of discretized differential
equations each time the cell is crossed by a packet.
In addition to the impact of photo-ionization and photo-heating, that, as
discussed above, are included in the form of discrete contributions, the
implemented chemistry network explicitly includes the other relevant
radiative processes: recombination and collisional ionization for the
ionization fractions and various cooling processes for the temperature
(bremsstrahlung, Compton cooling/heating, collisional ionization
cooling, collisional excitation cooling and recombination cooling).
All these radiative processes are instead implemented in the code as
{\it continuous}, namely by integrating their associated rate
coefficients.

We again refer the reader to the documentation mentioned above for
further details on the integration technique and on the technical
implementation devised to treat multiple sources.

\begin{figure*}
\centerline{\psfig{figure=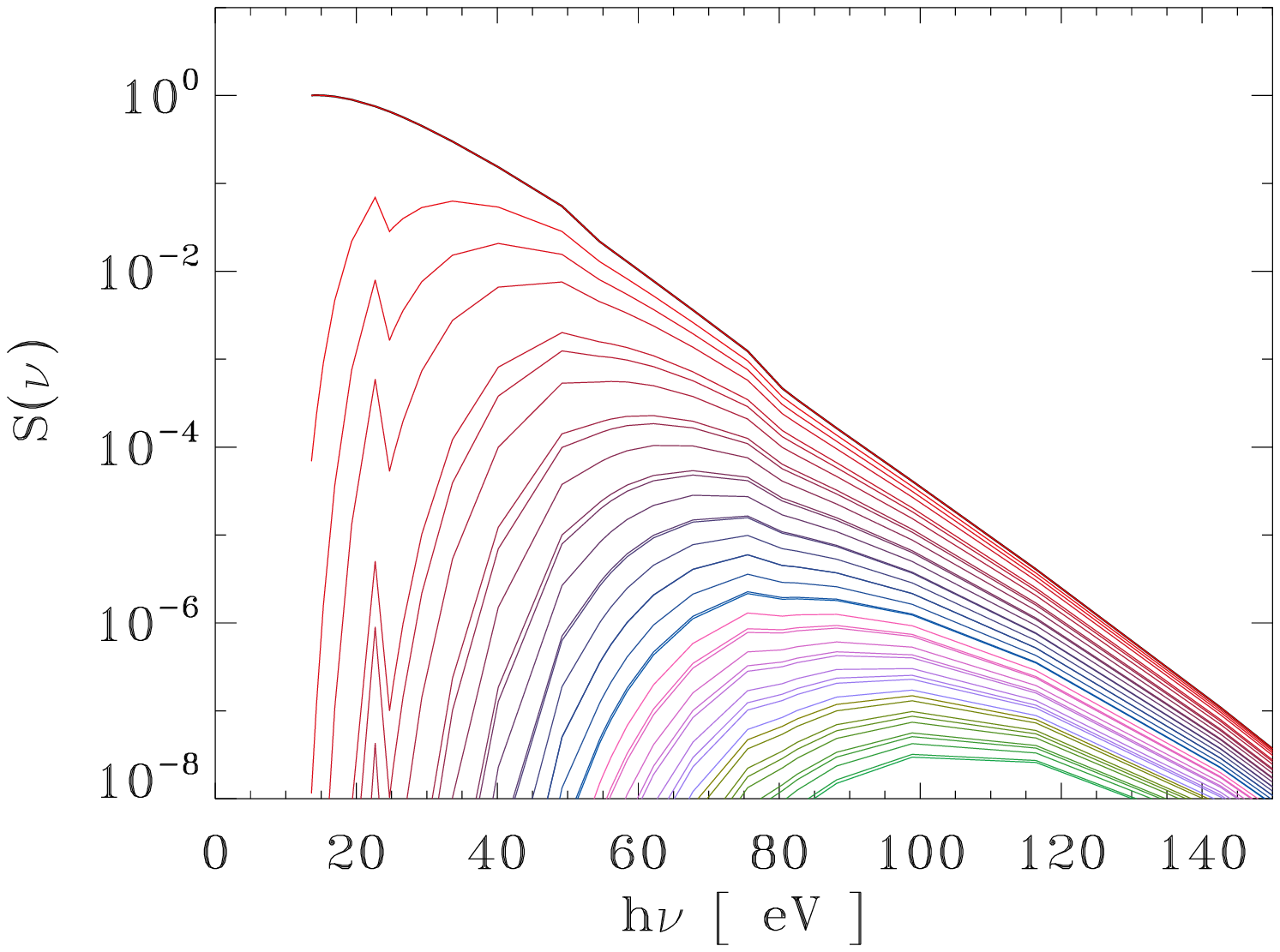,height=8.cm}\psfig{figure=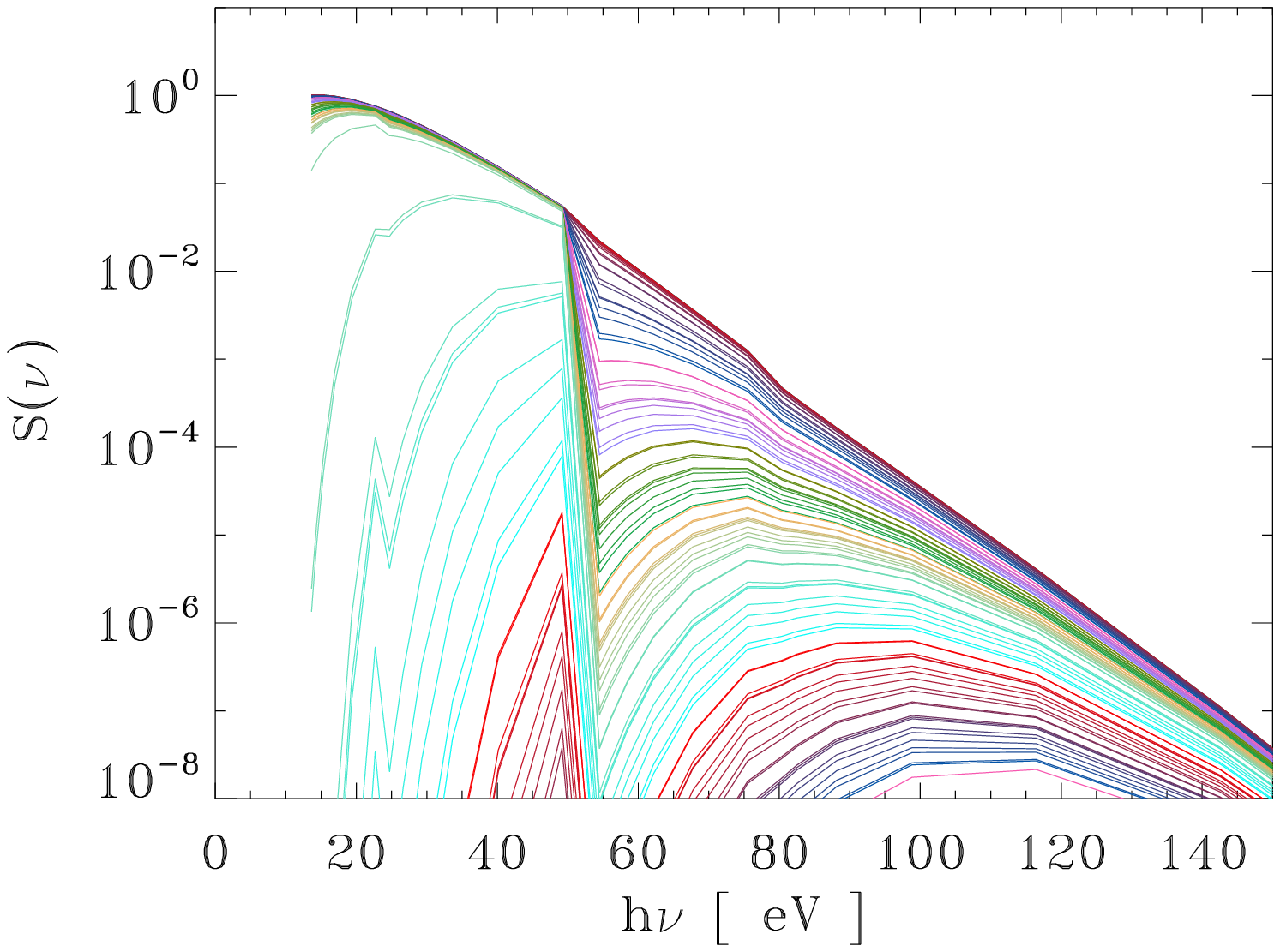,height=8.cm}}
\caption{
The curves show the evolution of a photon packet's
spectrum as the packet travels through the simulation volume, in a
H/He gas. From top to bottom, each curve corresponds to the spectrum 
at each cell in the crossing order. 
{\it Left}: The packet is emitted in a completely
neutral gas. {\it Right}:  The packet is emitted when the \HII / \HeII 
ionization fronts are roughly thirty cells away from the source.
}
\label{spectrum_distorsion}
\end{figure*}

\section{What is new in CRASH}
\subsection{Colored Packet}
In this section we describe the new implementation (in the following
we will refer to it as \crn)
developed to deal with ionizing sources with arbitrary spectra.
In {\tt CRASH1}, each point and diffuse ionizing
source is assigned with an arbitrary spectrum which is reproduced
by MC sampling the given spectral energy distribution (SED) and 
by emitting a large number of monochromatic photon-packets. 
This implementation is
highly versatile, allowing to reproduce arbitrary spectral
shapes in a continuum range of frequency, but has two main drawbacks
that could affect the precision and efficiency of the code.
First, the nature of the MC sampling procedure induces a poor sampling
of the spectral region with lower intensity, which in most
applications of interest is the hard
energy tail of the spectrum. This limitation could be crucial in
studies involving \eg helium (\HeII) photo-ionization, hard radiation
photo-heating (X-rays), ionization front (IF) profiles of hard
sources like quasars, just to give some examples among many.
Second, the monochromatic packet approach aggravates the well known
shortcoming of ray-tracing techniques, which oversample the regions
close the sources at the expense of undersampling the farthest ones.
This introduces unphysical features in the RT results, such as spikes
and uneven ionization fronts (\eg see Fig.~12 in I06).

To overcome these problems, we have introduced a new algorithm in \CR~
which treats the polychromatic nature of the ionizing radiation field
by shooting ``colored'' packets. Each packet is now represented
by a vector of $N_\nu$ bins, whose total energy $\Delta E$
(determined as in \crf) is distributed among a number of photons which
populates the $N_\nu$ frequency bins to reproduce the SED
characteristic of the ionizing source. The number and spacing of bins
can be arbitrarily chosen together with the frequency
range covered. Typically, the spacing is chosen to be logarithmic and more refined at the
photo-ionization thresholds, \ie $\nu_{\rm HI}$, $\nu_{\rm HeI}$ and $\nu_{\rm HeII}$.
Although we now have polychromatic photon packets traveling
through the box, their interaction with the gas has
not been modified significantly with respect to \crf.

While propagating through the gas the colored packet sees different
optical depths, one for each frequency bin:

\beq
\label{tau}
 \tau_c(\nu_j) = \sum_{i = {\rm HI,HeI,HeII}} \tau_i(\nu_j) = \sum_{i = {\rm HI,HeI,HeII}}  \delta l \times
\sigma_i(\nu_j) \times n_i,
\eeq

where the indexes $i$ and $j$ span over the species included and the frequencies
sampled, respectively. The other symbols in eq. \ref{tau} have the conventional
meaning, $\sigma_i$ and $n_i$ being the photo-ionization cross section
and the number density of the $i$ specie, and $\delta l$ the
actual path that the packet travels inside the cell. Differently from
the \crf~ implementation, here we have introduced a new algorithm which
computes the length $\delta l$ of the ray segment within each cell (see
next section).
Once the opacity of the cell is determined, the number of photons
deposited is calculated according to the probability for a photon to be
absorbed, $P(\nu_i)=(1-e^{-\tau_c(\nu_i)})$,
which depends on the atomic species and the frequency.

Summarizing, a packet is emitted from a source with
$N^0_\gamma(\nu_i)$ photons per frequency bin at the time $t_p$.
While traveling through the gas, photons are progressively
removed from the packet as they ionize atoms along the path. The number
of photons deposited in a cell is evaluated from the total optical
depth, which gives the probability for an absorption event:
\beq
\label{nphot}
N_{A,\gamma}^{(c)} (\nu_j)= N_{T,\gamma}^{(c-1)}(\nu_j) \left( 1 -
e^{-\tau_c(\nu_j)}\right),
\eeq
where $c$ counts the number of cells crossed by the packet from its
emission location, $N_{A,\gamma}$ the number of photons absorbed and
$N_{T,\gamma}$ the number of photons which is transmitted to the next
cell.

The absorption probability in the various frequency bins
induces a modification in the spectral shape of the packet. Typically,
the frequencies closer to the resonant photo-ionization
thresholds are depleted first and more strongly. 
In Figure~\ref{spectrum_distorsion} we show an example of how the
population of the frequency bins changes while the packet travels 
through a H/He gas. The differently colored curves represent the
spectral shape of the packet in the different cells pierced, with the 
sequence from top to bottom corresponding to the order in which they 
are crossed. 
Note that for both panels each color represents the cell after a given
number of crossings, \eg the transitions dark blue-to-magenta and
purple-to-green roughly correspond to the 20th and 30th cell crossed.
The two panels show the evolution of the spectral shape of a packet
emitted {\it (i)} in a completely neutral gas (left) and 
{\it (ii)} when the \HII / \HeII ionization fronts are roughly thirty
cells away from the source (right)\footnote{The physical configuration of the
simulation analyzed here is the same of the \CR/{\tt CLOUDY} comparison
described in Sec. 4.3, and the two panels correspond to packets
emitted at the simulation times $t_s = 0$ yr (left) and 
$t_s = 6 \times 10^4$ yr (right).}. 
It can be seen how the photons close to the photo-ionization
thresholds are absorbed first, due to the higher cross sections, and
how the SED of the packet becomes progressively harder as it
moves far away from the source.
It is interesting to notice that the spectrum evolution shown in the
left panel does not exhibit the drop at 54.4 eV, as here the packet is 
emitted in a completely neutral gas, still devoid of \HeII absorbers. 

A proper modeling of this filtering mechanism is very important,
particularly to correctly estimate the gas temperature or 
the inner structure of the ionization fronts.

Once the total number of photons deposited in a cell from each
frequency bin $\nu_j$ has been determined, $N_{A,\gamma}(\nu_j)$,
they are distributed
among the absorber species included in the calculation.
This is done by assigning to each absorber species a fraction of
$N_{A,\gamma}(\nu_j)$ proportional to the corresponding absorption
probability
\beq
\label{prob_s}
 P_i(\nu_j) \left(1-e^{-\tau_i(\nu_j)}\right),
\eeq
normalized to the total probability:
\beq
\label{prob_tot}
P_c(\nu_j) = \sum_{i = {\rm HI,HeI,HeII}}  P_i(\nu_j).
\eeq
As the packet passes through the cell, it photo-ionizes a number of
atoms of the $i$-th species given by:
\beq
N_{A_i,\gamma}(\nu_j)=N_{A,\gamma}(\nu_j)\times
\frac{P_i(\nu_j)}{P_c(\nu_j)},
\eeq
with a corresponding increment in the ionization fractions
given by:
\beq
\label{fracs}
\Delta x_{i^+} = \frac{\sum_j N_{A_i,\gamma}(\nu_j)}{N_{ii}},
\eeq
where $i^+$ is the singly ionized ion of the $i$ specie
($i^+ \in$ \{\HII, \HeII,\HeIII\} for $i \in$\{\HI, \HeI,\HeII\}), and
$N_{ii}$ the total number of nuclei of a given element contained in
the actual cell.

The cell temperature before the current packet crossing, $T_c$,
is updated accounting for photo-heating and for the
variation in the number of free particle which is given by the
variation in the number density of free electrons $n_e$. 
The two effects contribute respectively with the following 
temperature increments:
\be
\Delta T_{i^+} &= &\frac{2}{3k_Bn_c} \frac{\sum_j N_{A_i,\gamma}
(h\nu_j-h\nu_{th,i})}{V_{cell}}, \\
\Delta T_{n} &=& - \frac{2}{3} T_c \frac{\Delta n_e}{n_c},
\label{delta_temp}
\ee
where $h\nu_{th,i}$ is the ionization threshold for photoionization of 
the $i$-th specie, $n_c$ is the number density of free particles and 
$\Delta n_e$ is the variation of this quantity corresponding to the
change in the electron number density due to photo-ionization:
\be
\label{Delta_n}
\frac{\Delta n_e}{n_c}=\Delta x_e=\left[ f_h \times \Delta x_{\rm HII} +
f_{he} \times (\Delta x_{\rm HeII} + 2 \Delta x_{\rm HeIII})\right].
\ee

Each time a colored packet crosses a new cell the set of vectorial
operations described in equation \ref{tau} through \ref{Delta_n}
is performed to get the key quantities: the number of photons which escape
the $c$-th cell crossed from the emission location, 
$N_{T,\gamma}^{(c)}(\nu_j)= N_{T,\gamma}^{(c-1)}-N_{A,\gamma}^{(c)}$,
which accounts for radiative transfer, and $\Delta x_{i^+}$,
$\Delta T_{i^+}$ and $\Delta T_n$ necessary to account for the evolution of
the gas physical state.

The radiative processes which are not associated with the ionizing
radiation are reproduced, as done in \crf, by integrating the corresponding
temperature dependent rates over the elapsed time interval between 
the previous packet crossings and the current one.

\subsection{Ray Casting}

To limit the computational time, in the previous versions of \CR~ ray
casting is not implemented. Rather, the opacity of a single cell is
evaluated using as the length traveled by the packet within the cell
the fix value 0.56 (in units of the linear dimension of a cell). This
number corresponds to the median value of the PDF for the length of
randomly oriented paths within a cubic box of unit size (see Ciardi
\etal 2001).

Although this approximation is acceptable for several applications,
it does not always result in enough accuracy, occasionally showing
spurious geometrical patterns in the ionization and temperature maps
(see the rings evident in the bottom right panel of Figure~\ref{T2_maps}).
For this reason, a new algorithm for ray casting has been
implemented, which is similar to the {\it Fax Voxel Trasversal
  Algorithm for Ray Tracing} by Amanatides \& Woo (1987). 
Ray casting is performed by computing the distances that need to be 
traveled in order to intersect the boundaries (along the $x$, $y$ and $z$ axis)
of the grid cell pierced by the ray. 
The minimum of these distances gives the length of the path traveled 
within the cell and the boundary intersected first determines which 
index has to be incremented to get the coordinate of the next cell 
through which the ray passes. 
For more technical details about the algorithm we refer the reader 
to the Amanatides \& Woo (1987) paper. \\
The test cases described in Sec.~4 show the improvement resulting
from the introduction in \crn~of the ray casting calculation described above.
This improvement in performance comes at the expense of
computational efficiency, which nevertheless is not dramatically
affected. The fraction of total computational time spent in
computing ray casting strongly depends on the specifics of a given
run, particularly on the physics included. In all the cases analyzed here
we find it to be almost negligible, being at most five percent of the 
total computational time.

\subsection{Tables and Rates}

In the new version of the code we have updated the expressions
adopted for the coefficient rates. The references for each specific 
coefficient can be found in Table~1 of I06. 
We have furthermore introduced the use of pre-compiled tables 
for the rate coefficients with the aim of improving efficiency. 
The temperature dependence of the various coefficient rates are now 
computed once at the beginning of a given run, stored in dedicated 
tables and read during the run. This allows to speed up the
calculation significantly with respect to the previous implementation 
where the coefficients rates are functions evaluated
at each cell crossing. 
Although a quantitative estimate of the improvement in the efficiency
could be given, this is highly dependent on the problem at hand and would
be significant only in specific cases. Typically though, we find that
the efficiency is increased up to $\sim$~20 percent.
The implementation is worked out to allow for adaptive tables, 
by assigning as initial conditions the range of temperatures and the
refinement of sampling according to the physical configuration 
associated to a given run. This is done once, at the beginning of each
run, with the aim of adapting the range and refinement of the tables to
the physical configuration of the simulation at hand: \eg if the gas 
in the simulation has been shock heated we assign an upper limit 
on the temperature range of the order of $T_{max} \approx 10^7$K, 
which is not the case if we know that the run will process only 
photoionized gas, for which we adopt $T_{max} \approx 10^5$K.

\section{Results and Performance}
In the following we will discuss some test cases run to study the performance 
of the new implementation described above.

\begin{figure*}
\centerline{\psfig{figure=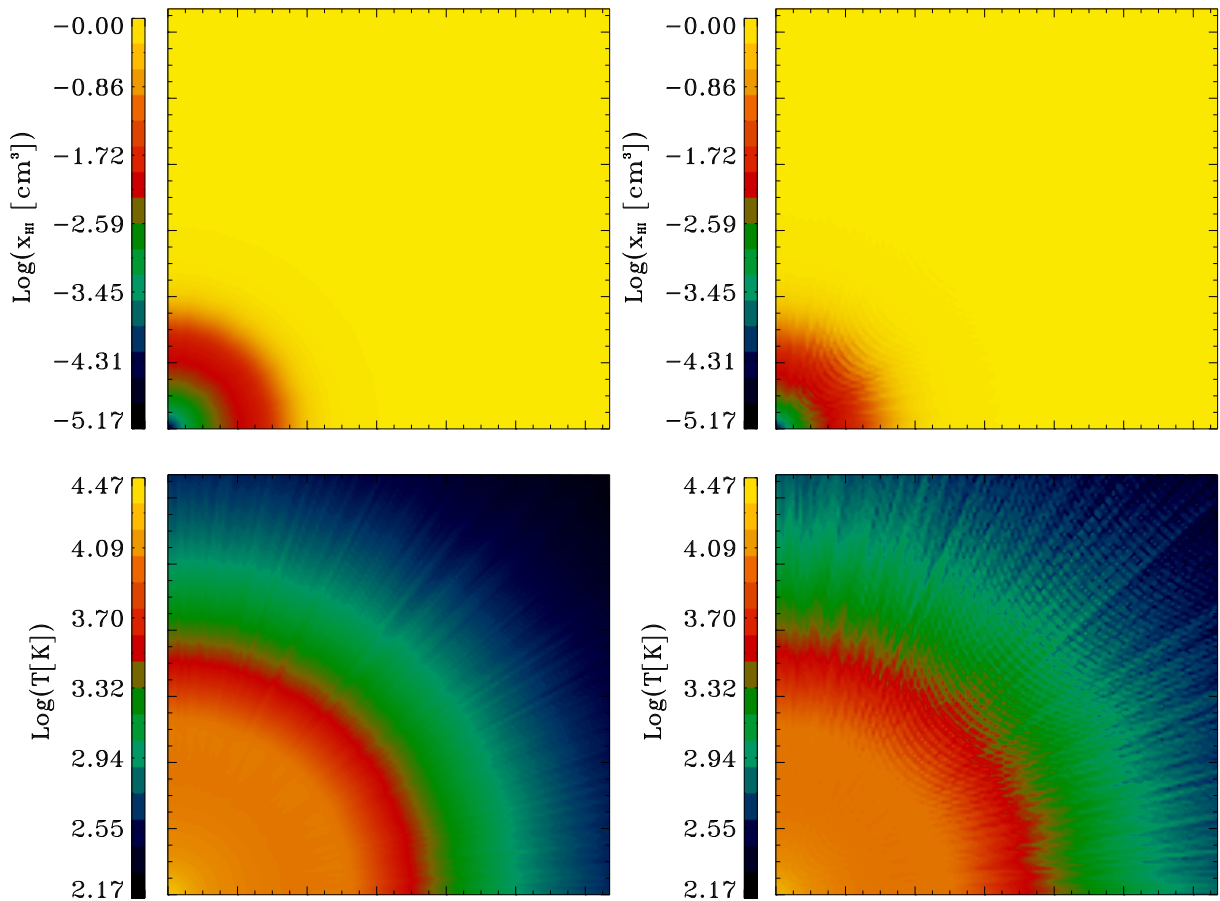,height=13.cm}}
\caption{Test~2 of the RT Comparison Project (\HII region expansion in a
uniform H gas with varying temperature). The upper (lower) panels show
maps of the \HI fraction (temperature) cut through the simulation volume
at coordinate $z=0$ and time $t=10$~Myr. {\it Left}: \crn;
{\it Right}: \crf.}
\label{T2_maps}
\end{figure*}

\subsection{Comparison Project Test~2}

As a first test we have repeated Test 2 of the RT Comparison Project
(I06). This test is the classical problem of an \HII
region expansion in a uniform gas around a single ionizing source.
A steady source with a $10^5$~K black-body spectrum, emitting
$\dot{N}_\gamma=5 \times 10^{48}$ ionizing photons per unit time ($s^{-1}$), 
is turned on in an initially-neutral, uniform-density, static
environment with hydrogen number density $n_H=10^{-3}$~cm$^{-3}$.
The computational box dimension is L = 6.6 kpc and the source is positioned
at its corner.
We follow the evolution of the neutral hydrogen fraction and of the 
gas temperature (initially set to 100~K), running the \crf~ version
with $10^9$ photon packets and the \crn~ one with $2\times 10^8$ 
photon packets. 

The \HI fraction and temperature maps, 10~Myr after the source has turned on,
are shown in Figure~\ref{T2_maps} for {\tt CRASH1} (right) and {\tt
 CRASH2} (left) (same cuts in Figs.~11 and 12 of I06).  
From a comparison between the left and right panels, differences are
evident. 
More specifically, the spikes present in the \crf~ 
maps, particularly in the temperature, are almost completely
vanished. This is due to the colored packets implementation
that allows for a much better sampling of the high energy tail of the spectrum,
for the same number of packets emitted (and thus for the same resolution).
The weak spikes-like features which are still visible in the
\crn~ temperature map, are due to the noise intrinsic to Monte Carlo algorithms.
Also, the artificial ring-like patterns clearly visible in the \crf~
maps, have completely disappeared thanks to the new ray casting algorithm implemented.
Finally a visual inspection of the maps in Figure~\ref{T2_maps}
reveals that the volume inside the I-fronts is both hotter and
more highly ionized in {\tt CRASH2} than in \crf.
\begin{figure*}
\centerline{\psfig{figure=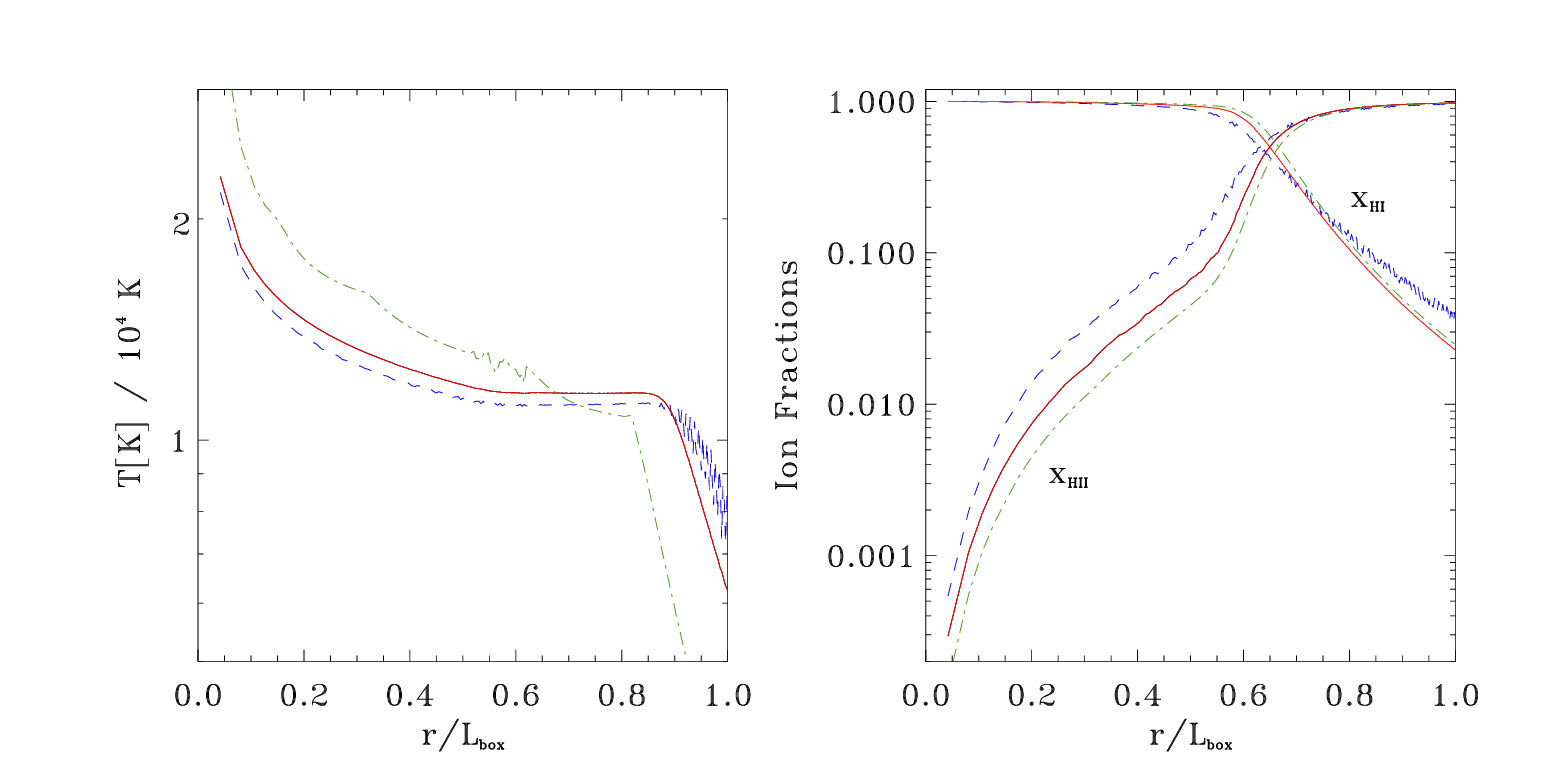,height=9.cm}}
\caption{Test~2 of the RT Comparison Project (\HII region expansion in an
uniform H gas with varying temperature). The left (right) panel shows the
spherically-averaged temperature (ionized and neutral H fraction) as a
function of the distance from the source, 100~Myr after the source has
turned on. {\it Solid line}: \crn; {\it Dashed line}: \crf; 
{\it Dashed-Dotted line}: {\tt C$^2$-RAY}.
}
\label{T2_prof}
\end{figure*}

This can be better seen in the more quantitative comparison shown in
Figure~\ref{T2_prof}, which plots the spherically-averaged temperature 
(left panel) and ionized and neutral fractions (right panel) as a function
of the distance from the source 100~Myr after the source has been
turned on. {\tt CRASH2} (solid red lines) produces a 
higher temperature and a lower neutral fraction compared to 
{\tt CRASH1} (dashed blue lines). This
result, which brings a better agreement between \CR~ and other codes (\eg
{\tt C$^2$-ray}, whose results are shown for comparison by the
dashed-dotted green curves), is due once again the improved modeling of
the filtering which is achieved thanks to a better sampling of the
high energy tail of the spectrum: photons in this region of the SED
can heat more and ionize less the gas with respect to lower energy
photons which are oversampled in \crf. Despite this, as a result of
the increased temperature which inhibits recombinations, the ionization
fraction within the ionized sphere results higher in \crn.
One can also see that the ionization front lies slightly closer to the 
ionizing source in the new implementation: this happens because, once
the luminosity is assigned as energy emitted in the unit time, the 
oversampling of lower energy part of the SED (which affects the 
\crf~ implementation) results in an higher number of ionizing photons.
Also, we still obtain a wider ionization front with respect
to other codes (see Fig.~17 in I06), which results from the better
sampling of the spectrum in the frequency space.
\begin{figure*}
\centerline{\psfig{figure=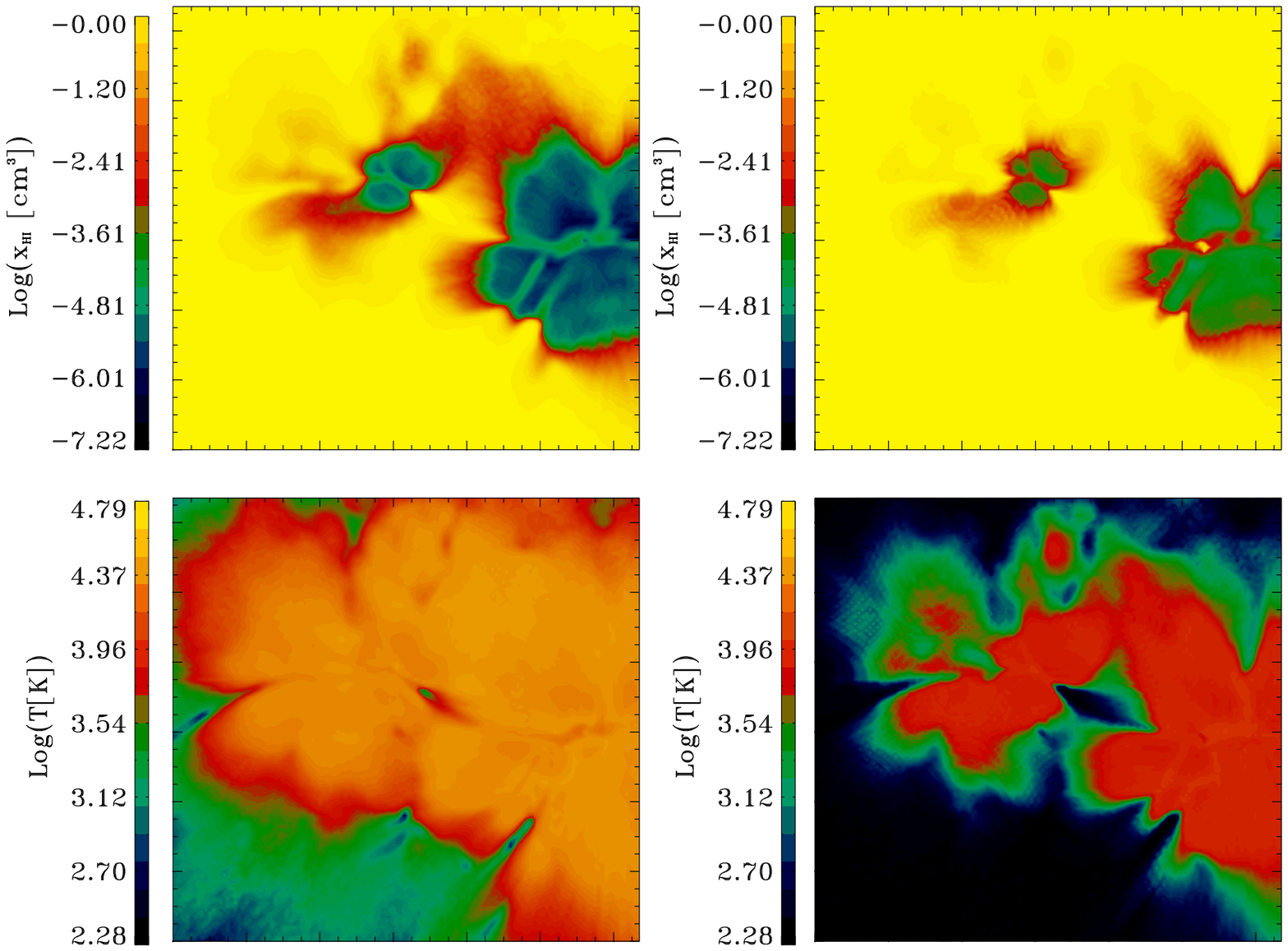,height=13.cm}}
\caption{Test~4 of the RT Comparison Project (reionization of a
cosmological density field). The upper (lower) panels show
maps of the \HI fraction (temperature) cut through the simulation volume
at coordinate $z=z_{box}/2$ and time $t=0.05$~Myr. {\it Left}: \crn;
{\it Right}: \crf.} 
\label{T4_maps}
\end{figure*}
\subsection{Comparison Project Test~4}
\begin{figure}
\centerline{\psfig{figure=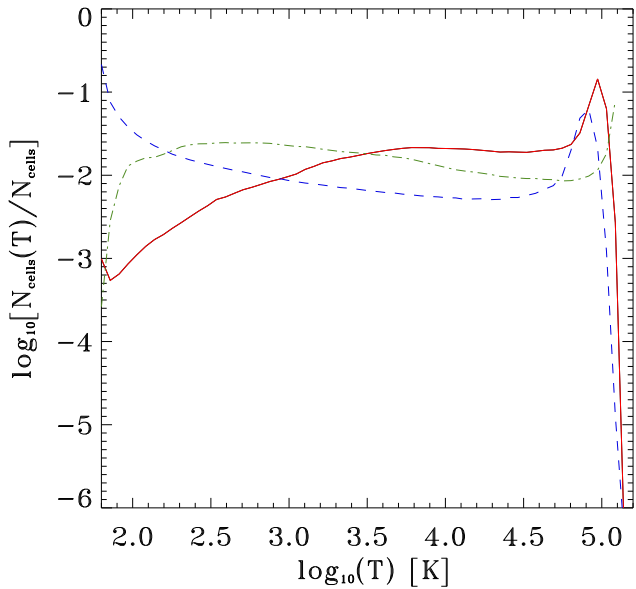,height=9.cm}}
\caption{
Test~4 of the RT Comparison Project (reionization of a
cosmological density field): histogram for the temperature at time
$t=0.05$~Myr. {\it Solid line}: \crn;
{\it Dashed line}: \crf; {\it Dashed-Dotted line}: {\tt C$^2$-RAY}.
}
\label{T4_hist}
\end{figure}

In this section we describe the results from the Test~4 of the RT
Comparison Project. In this case, the propagation of ionization fronts from
multiple sources in a static cosmological density field is followed.
The initial conditions are provided by a time-slice (at redshift $z=9$)
from a cosmological N-body and gas-dynamic simulation. The simulation
box size is 0.5$h^{-1}$ comoving Mpc. The gas is assumed to be
initially neutral and the temperature is initialized at 100~K everywhere in the box.
The ionizing sources are chosen so as to correspond to the 16 most
massive halos in the box, which are identified with a friends-of-friends
algorithm. We assume that the sources have a black-body
spectrum with effective temperature $T_{eff}=10^5$~K. The ionizing
photon production rate for each source is constant and is assigned
assuming that each source lives for 3~Myr and emits 250
ionizing photons per atom during its lifetime.
For simplicity, all sources are assumed to switch on at the same
time. The boundary conditions are transmissive (\ie photons leaving
the computational box are lost, rather than coming back in as for
periodic boundary conditions).
The maps in Figure~\ref{T4_maps} show the \HI fraction (upper panels) and temperature
(lower panels) cuts through the simulation volume at $z = z_{box}/2$
and at time 0.05 Myr (same cuts in Fig.~31 and Fig.~32 of I06).
Both {\tt CRASH1} (right panels) and {\tt CRASH2} (left panels)
outputs are shown in the figure.
The differences between the two implementations are strikingly
evident: as already found in the Test 2 above, the ionization fronts are
much larger in the \crn~ implementation which accounts properly for the  
higher energy photons associated with larger mean free paths. The
better sampling of the harder tail part of the spectrum results
also in higher temperature and lower neutral hydrogen fractions as
already discussed above.  
The new ray casting implementation corrects furthermore for the
artificial patterns visible close-by the ionization fronts in the \crf.
The difference in the internal structure of the ionized region can be
better appreciated in Figure~\ref{T4_hist} where the histograms of the
temperature at time $t=0.05$~Myr obtained with the two implementations
of the code are compared: solid (dotted) line is for {\tt CRASH2} ({\tt CRASH1}).
With the colored packets implementation we find on average higher
temperatures. The fact that regions at temperature above $10^3$~K
are more densely populated reflects the higher temperature of ionized
regions produced by the contribution of harder photon packets which 
are undersampled in \crf. 
At the same time the lack of the peak found at low temperatures in
\crf~  is due to the wider ionization fronts obtained with \crn, which
extend much further from the sources and affect almost the whole
computational volume. 
As in the previous test, with \crn~ we found a better agreement 
with {\tt C$^2$-ray} results shown in the figure by the dashed-dotted
curve\footnote{It should be noted that in Fig.~37 of
I06 the purple and green lines have been erroneously exchanged.}.\\
\subsection{CRASH/CLOUDY Comparison}
Finally, similarly to what was done in MFC03, we compare our code, with
the publicly available 1D RT code
{\tt CLOUDY94}\footnote{http://nimbus.pa.uky.edu/cloudy}. 
With this test we seek to check for the performance of the new algorithm in 
dealing with helium physics, which has not been included in the 
comparison project tests. 
We consider the case of a point source with a black body spectrum at
$T=6\times10^4$~K and luminosity $L=10^{38}$~erg s$^{-1}$, 
ionizing a uniform medium with density $n=1$~cm$^{-3}$ in a gas
composed by hydrogen (90 percent in number density) and helium. 
\begin{figure*}
\centerline{\psfig{figure=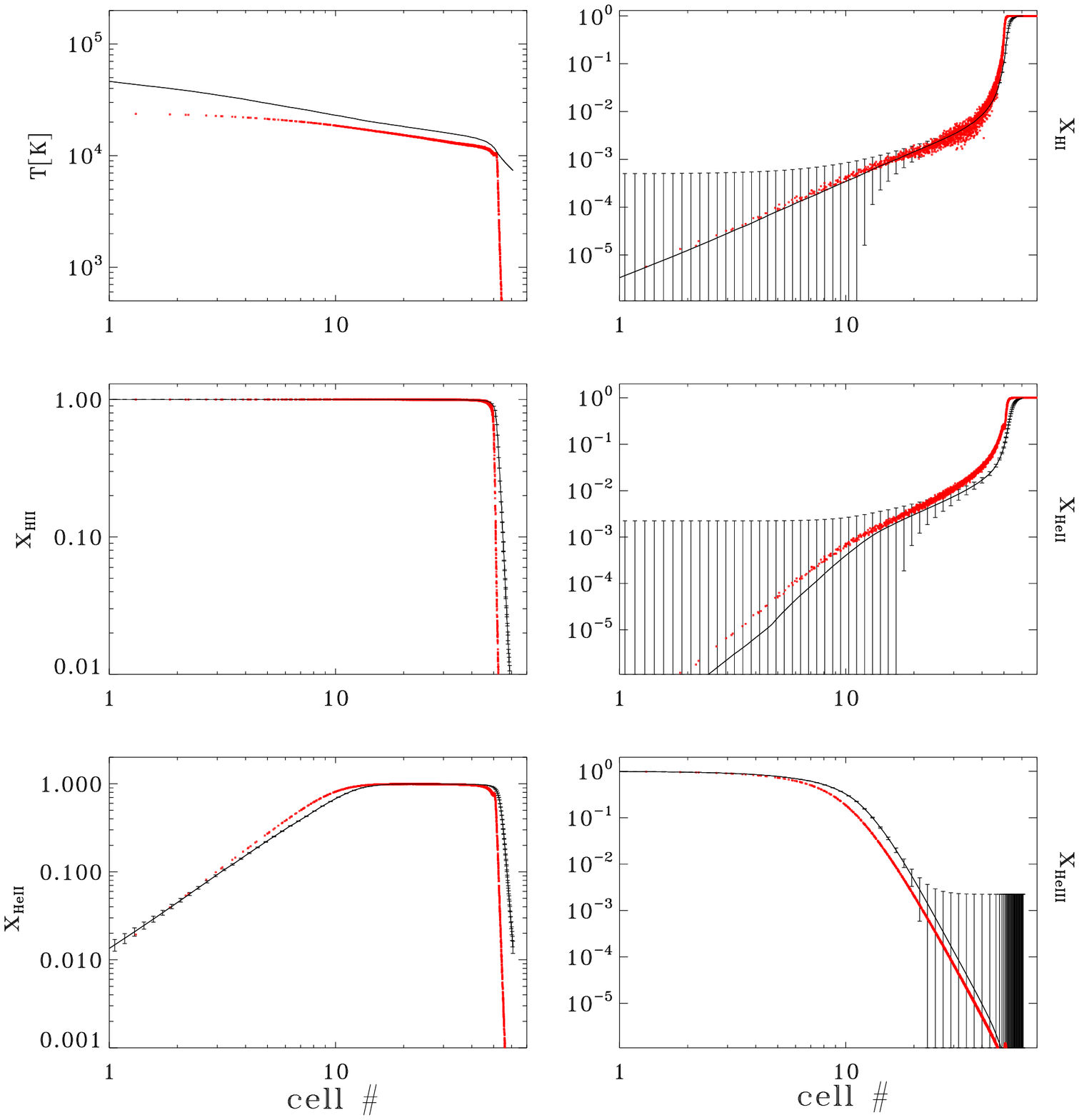,height=15.cm}}
\vskip 1truecm
\caption{
Comparison between the equilibrium configurations obtained by
\crn~ (points) and {\tt CLOUDY94} (solid lines)
for different physical quantities, as a function of distance 
from the point source in cell units.
The errors refer to the {\tt CLOUDY94} results and are evaluated
as explained in the text.
}
\label{C_C}
\end{figure*}
Figure~\ref{C_C} shows the profiles at equilibrium for temperature,
neutral and ionization fractions. Compared to the earlier
results of {\tt CRASH1} (see Fig.~4 in MFC03), the new version of the code
provides an overall much better agreement with {\tt CLOUDY94}.
In particular the neutral hydrogen fraction inner profile fits
extremely well and with reduced dispersion the {\tt CLOUDY94} predictions. 
The same is true for the neutral helium fraction profile which is however
slightly higher in \crn~ than in {\tt CLOUDY94}. 
Nevertheless, both the temperature and the \HeII profiles are
slightly lower than the ones derived with {\tt CLOUDY94} and the
ionization front extends somewhat further in {\tt CLOUDY94}.
The two behaviors are likely to be correlated and might be due to differences in
the expressions adopted for the coefficient rates associated to helium
together with the fact that \CR~ neglects the effects of heat conduction.

\section{Summary}
In this paper we introduce the new version of the radiative
transfer code \CR. The details of the code implementation have been
previously described in MFC03. Here we present the new elements
introduced to improve performance and efficiency.
The main new ingredients implemented in \crn~ are: 
\begin{itemize}
\item the introduction of colored photon packets, which brings 
a significant improvement in the sampling  of source spectra both in
space and frequency; 
\item a new model for propagating the photon packets
through the Cartesian grid, which includes a proper ray casting
algorithm; 
\item the use of pre-compiled adaptive tables for the
coefficient rates which result in more efficient computing. 
\end{itemize}
The \crn~ implementation has been tested and compared with the earlier 
versions of the code by performing a set of reference
tests taken from the RT codes comparison project (I06) and from
MFC03. \crn~ is more accurate both in the calculation of the
temperature and the ionization fractions, thanks to the improved sampling
of the hard tail part of ionizing sources spectra, which is undersampled in the
monochromatic packet implementations. Furthermore, the
better sampling of hard radiation together with the new ray casting
algorithm corrects for spurious geometrical features like spikes in
Str\"omgren sphere boundaries or artificial geometrical pattern that
are found in the \crf~ results. 

\section*{Acknowledgments}
We thank the anonimous referee for her/his helpful comments on the draft. 
We are grateful to Patrik Jonsson and Adrian Partl for useful discussions.
We thank Marco Pierleoni for his sharp comments on the technical
details of the implementation.
AM is supported by the DFG Priority Program 1177.

\label{lastpage}
\end{document}